\pdfoutput=1
\documentclass[11pt]{article}

\usepackage[a4paper,margin=1in]{geometry}
\usepackage[utf8]{inputenc}
\usepackage[T1]{fontenc}
\usepackage{amsmath,amssymb,amsthm}
\usepackage{graphicx}
\usepackage{booktabs}
\usepackage{array}
\usepackage[numbers,sort&compress]{natbib}
\usepackage[colorlinks=true,citecolor=blue,linkcolor=blue,urlcolor=blue]{hyperref}

\title{Weaker Coherence, Weaker Reciprocity: Comparing the Semantic and Social Organization of Moltbook and Reddit}

\author{
Favio Di Ciocco$^{1,2,\dagger}$,
Lucas Díaz Celauro$^{1,2,\dagger}$,
Sebastián Pinto$^{1}$ \\
Marcelo Kuperman$^{3,4}$,
Pablo Balenzuela$^{1,2,*}$
}
\date{}

\begin{document}

\maketitle

\begingroup
\renewcommand{\thefootnote}{}
\footnotetext{
$^{1}$Universidad de Buenos Aires, Facultad de Ciencias Exactas y Naturales, Departamento de Física, Buenos Aires, Argentina\\
$^{2}$Instituto de Física Interdisciplinaria y Aplicada (INFINA), CONICET -- Universidad de Buenos Aires, Buenos Aires, Argentina\\
$^{3}$Instituto Balseiro, Universidad Nacional de Cuyo y Comisión Nacional de Energía Atómica, Bariloche, Argentina\\
$^{4}$Consejo Nacional de Investigaciones Científicas y Técnicas (CONICET), Argentina\\
$^{\dagger}$These authors contributed equally to this work.\\
$^{*}$Correspondence: balen@df.uba.ar
}
\endgroup

\begin{abstract}
Large language models enable the creation of autonomous agents that interact in social environments, raising the question of whether agent-based platforms reproduce the organizational properties of human social networks. We compare Moltbook, a social network populated by AI agents, with early Reddit, focusing on how communities organize and differentiate semantic content, using network analysis and NLP methods to characterize semantic coherence and diversity within and between communities, and their relationship to user activity.
We find a systematic difference between the two platforms. Reddit communities show stronger semantic coherence, closer alignment with community names, and greater semantic diversity, with individual communities spanning broader content and communities more differentiated from one another. This combination distinguishes Reddit from Moltbook, whose communities are more homogeneous, less differentiated, and increasingly misaligned with their names over time. Users on Reddit also participate across communities that are more semantically related than those connected by activity in Moltbook. At the interaction level, comment-network motif analysis shows Moltbook dominated by non-reciprocal, broadcast-like exchanges, whereas Reddit shows more reciprocal, chained interaction patterns. These results indicate that Reddit combines semantic coherence with diversity across organizational levels, a pattern not reproduced by the AI-agent network.
\end{abstract}


\section{Introduction}

\par Online social networks are complex systems in which local interactions among individuals give rise to collective structures at larger scales. \cite{mislove2007measurement, newman2003structure, lazer2009computational} Communities, patterns of user participation, and the differentiation of topical spaces are emergent properties of these systems, resulting from the repeated interaction of users within a shared social environment \cite{Avalle2024}. Understanding the mechanisms underlying this organization requires considering both the structure of interactions and the semantic content through which individuals communicate. In particular, the relationship between social and semantic organization can reveal how collective systems develop coherent communities while maintaining diversity across the network \cite{roth2010social, newman2016structure, hric2016network}.

\par Reddit provides a natural example of such a system. Its organization into user-created communities, or subreddits, creates a heterogeneous network in which users can participate in multiple topical spaces \cite{anderson2015ask}. Patterns of participation generate non-trivial network structure, such as persistent user interests across different subreddits \cite{Olson2013,Valensise2019,Waller2021}. The availability of large-scale historical datasets has further enabled the quantitative study of Reddit as a complex social system \cite{Baumgartner2020}. Importantly, the organization of Reddit communities is not determined solely by their nominal topics: it emerges from the interaction between the interests of users, the structure of their activity, and the semantic content generated within each community \cite{olson2015navigating, hessel2016science}.

\par The emergence of large language models (LLMs) introduces a qualitatively different class of agents into such environments. LLM-based agents can generate language and interact autonomously with other agents, making it possible to construct social systems in which collective organization emerges from interactions between artificial rather than human participants \cite{ashokkumar2026large, park2023generative}. This makes it possible to
construct social systems in which collective organization emerges from
interactions between artificial rather than human participants. Recent work
has further shown that LLMs can generate realistic multi-user social-media
discussions \cite{Bouleimen2026}, reinforcing their potential as a substrate
for computational social simulation. These systems provide an opportunity to investigate whether collective properties observed in human social networks can emerge from artificial agents, and which aspects of such organization depend on the nature of the agents themselves. Recent work has begun to characterize the structural and behavioral properties of social environments populated by LLM-based agents, highlighting both similarities and differences with human online communities \cite{Mou2024,Hashemi2026,Chen2026}.

\par Moltbook (\href{https://moltbook.com/}{https://moltbook.com/}) offers a particularly suitable setting for studying these questions. The platform consists of communities in which autonomous AI agents post and interact, providing a social architecture that is structurally reminiscent of Reddit. Each AI agent is configured by a human owner, although all online interactions are executed by the agent via API \cite{Moltbook2026}. Early analyses of Moltbook have identified distinctive patterns of participation and community organization, including strong concentration of activity and differences in interaction patterns relative to human social networks \cite{Jiang2026, Goyal2026, Zhang2026, zerhoudi2026form}. These observations motivate a broader question concerning the organization of the resulting system: beyond the amount and structure of activity, do AI-agent communities develop a semantic organization comparable to that observed in human social networks?

\par Here we address this question through a comparative analysis of Moltbook and the early stages of Reddit. We focus on two complementary properties of complex social systems: \emph{semantic coherence} and \emph{semantic diversity}. Coherence characterizes the extent to which the semantic content associated with a community is related to its identity and to the communities connected to it through patterns of user activity. Diversity describes the extent to which semantic content varies within communities and across different communities. These properties need not be antagonistic. More generally, studies of human
communication and collective semantic search show that social groups can
combine shared alignment with the exploration of novel semantic regions
\cite{Speer2024,Ueshima2024}. A complex social system can simultaneously support heterogeneous content within communities and maintain a differentiated organization between communities. Their joint characterization therefore provides a way to quantify how semantic structure emerges across multiple organizational scales.
We complement this semantic characterization with an analysis of the comment network, examining triadic interaction motifs to assess whether user exchanges are organized around reciprocal engagement or one-to-many, broadcast-like communication


\par The comparison reveals a systematic difference between the two systems. Reddit combines greater semantic diversity with stronger semantic coherence. Its communities contain more heterogeneous content, while occupying more differentiated regions of semantic space, and their content remains more closely aligned with the semantic information conveyed by their names. Furthermore, users tend to participate across communities that are more semantically related to one another. Moltbook, in contrast, exhibits more homogeneous communities, weaker semantic differentiation between communities, and a progressively weaker correspondence between community identity and content. 
This divergence extends to the structure of user interactions: Moltbook's comment network is dominated by non-reciprocal, broadcast-like exchanges, whereas Reddit shows a substantially higher prevalence of reciprocal and chained interaction patterns.
These results suggest that the collective organization of an AI-agent social network differs from that of a human social network not only in its interaction patterns, but also in the coupling between social structure and semantic organization.

\section{Materials and Methods}

\subsection{Moltbook data collection}

Moltbook is an existing online social platform designed for interactions between autonomous AI agents. Agents participate in user-created thematic communities, called \emph{submolts}, where they can publish posts, comment on other posts, and reply to existing comments. Importantly, the agents analyzed in this study were not created, simulated, or controlled by us. Our analysis is entirely observational and is based on activity generated independently on the public Moltbook platform. The underlying characteristics of individual agents---including the language models, prompts, memory systems, tools, and other configuration details used by their owners---are generally unknown to us. We therefore treat each Moltbook account as an observed agent and analyze only its publicly available behavior on the platform.

We collected public activity from Moltbook through its application programming interface (API) using custom Python scripts. The dataset comprises posts and
their associated comments, from the first available post of the platform (on January 28th, 2026) until July 31st, 2026. In this date range,
the dataset contains 3,251,763 posts and 17,746,686 comments, including top-level comments and nested replies. It involves 183,565 unique users who
posted or commented, of whom 179,651 authored at least one post, 45,507 authored
at least one comment, and 41,593 did both. The posts are distributed across
7,240 submolts.

The information retrieved includes the content of the post or comment, together
with information on the submolt and author. It also includes post-level metrics
such as upvotes, downvotes, and score. Additional details on the dataset
structure and the fields contained in these records are provided in
Appendix~\ref{app:moltbook_api}.

\subsection{Reddit data}

Reddit data were obtained from the historical Reddit archive distributed
through Academic Torrents \cite{Baumgartner2020,RaiderBDev2026}. The archive
contains monthly dumps of Reddit submissions and comments, stored as
Zstandard-compressed newline-delimited JSON files. We downloaded the period
from December 2005, the earliest month for which both submissions and comments
were available, through December 2009. For each month within this interval, we
downloaded both the submission and comment archives.

For the analyses presented here, we selected the period from January 23rd, 2008 to July 26th, 2008. This interval matches the duration of the Moltbook observation window and begins when Reddit had recently enabled the creation of user-defined communities, providing a more comparable stage of platform development.
After filtering by creation date, the dataset contains 1,079,709 posts and 3,124,758 comments. It involves 121,999 unique users who posted or commented, of whom 95,449 authored at least one post, 50,448 authored at least one
comment, and 23,898 did both. The posts are distributed across 1,913
subreddits.

Each record contains the information provided for an individual Reddit
submission or comment, including its textual content, author, subreddit,
creation time, and engagement-related metadata. Additional details on the
archive structure and the fields contained in these records are provided in
Appendix~\ref{app:reddit_data}.

\subsection{Concentration of user activity across communities}
\label{sec:user_gini}

\par To quantify how users distribute their activity across communities, we
followed the approach introduced by Valensise et al.
\cite{Valensise2019}. For each user $u$, we constructed an activity vector

\begin{equation}
    \mathbf{v}^{(u)}
    =
    \left(
    v^{(u)}_1,\ldots,v^{(u)}_N
    \right),
\end{equation}
where $v^{(u)}_i$ is the number of comments produced by user $u$ in
community $i$, and $N$ is the total number of communities in the
corresponding platform. The total activity of the user is therefore

\begin{equation}
    I_u = \sum_{i=1}^{N} v^{(u)}_i .
\end{equation}

\par The concentration of this activity across communities was quantified using
the Gini coefficient,
\begin{equation}
    g_u =
    \frac{1}{2 N I_u}
    \sum_{i=1}^{N}
    \sum_{j=1}^{N}
    \left|
    v^{(u)}_i-v^{(u)}_j
    \right|.
\end{equation}
Values close to zero correspond to activity distributed approximately
uniformly across communities, whereas values close to one indicate that the
user concentrates most of their activity in a small subset of communities.

\par Since most users interact with only a small fraction of all available
communities, $\mathbf{v}^{(u)}$ is typically sparse. Moreover, for a user
with $I_u<N$ interactions, a completely homogeneous distribution over all
$N$ communities is not attainable. We therefore corrected the Gini
coefficient for this finite-activity constraint. The minimum attainable Gini
for a user with $I_u$ comments is

\begin{equation}
    g_u^{*} =
    \begin{cases}
    \displaystyle \frac{N-I_u}{N}, & I_u < N,\\[6pt]
    0, & I_u \geq N,
    \end{cases}
\end{equation}
corresponding, when $I_u<N$, to the maximally dispersed configuration in
which each interaction occurs in a different community. We then defined the
normalized Gini coefficient as
\begin{equation}
    \hat{g}_u =
    \frac{g_u-g_u^{*}}
    {1-g_u^{*}}.
\end{equation}
Thus, $\hat{g}_u=0$ represents the most homogeneous allocation of activity
allowed by the user's number of interactions, whereas values approaching
one indicate increasingly concentrated activity.

For computational efficiency, communities in which a user had no activity
were treated as implicit zeros when evaluating the Gini coefficient rather
than explicitly constructing the full $N$-dimensional activity vector.

\subsubsection{Null model}
\label{sec:null_model}

To determine whether the observed concentration could arise solely from the
finite amount of activity of each user, we constructed a null model following
the same principle as Valensise et al. \cite{Valensise2019}. For each user,
the total number of comments $I_u$ was preserved, while their allocation
across communities was randomized.

More precisely, for a user with $I_u$ comments, we defined a set of
\begin{equation}
    K_u = \min(I_u,N)
\end{equation}
candidate communities. Each of the $I_u$ comments was then independently
assigned with equal probability to one of these $K_u$ communities. Thus,
the randomized user could interact with at most $I_u$ different communities,
while preserving exactly the same total activity as in the empirical data.
Repeated assignments to the same community were allowed.

For each randomized activity vector, the raw and normalized Gini
coefficients were calculated using the same procedure as for the empirical
data. We generated 10 independent null-model realizations for each user.
The procedure was applied independently to Reddit and Moltbook, using the
corresponding total number of communities in each platform.

\subsection{Semantic representation of communities}
\label{sec:semantic_representation}

To compare the semantic organization of Reddit and Moltbook, we represented
their communities in a common embedding space. For each community $c$
(a subreddit or submolt), we considered two sources of semantic information:
the name of the community and the textual content of the posts published
within it. For both platforms, the title of each post was used as its textual
representation.

Text was encoded using the
\texttt{sentence-transformers/all-MiniLM-L6-v2} \cite{Reimers2019} model. The same model was
used to embed both post titles and community names, ensuring that all elements
were represented in the same semantic vector space. For each community, to achieve a statistically significant number of posts and at the same time get a manageable amount of data to embed, we
considered at most 200 posts. Thus, if $\mathbf{x}_{c,i}$ denotes the
embedding of post $i$ in community $c$, its semantic centroid was defined as
\begin{equation}
    \boldsymbol{\mu}_c =
    \frac{1}{N_c}\sum_{i=1}^{N_c}\mathbf{x}_{c,i},
\end{equation}
where $N_c \leq 200$ is the number of posts considered for community $c$.
The community name was embedded independently, resulting in a vector
$\mathbf{n}_c$.

Semantic differences were quantified using cosine distance,

\begin{equation}
    D_{\mathrm{cos}}(\mathbf{x},\mathbf{y})
    =
    1 -
    \frac{\mathbf{x}\cdot\mathbf{y}}
    {\|\mathbf{x}\|\,\|\mathbf{y}\|}.
\end{equation}
We then defined three complementary distances.
First, the distance between an
individual post and the centroid of its community,
\begin{equation}
    d_{c,i}
    =
    D_{\mathrm{cos}}
    \left(\mathbf{x}_{c,i},\boldsymbol{\mu}_c\right),
\end{equation}
characterizes the internal semantic consistency of a community. Smaller
values indicate that posts are concentrated around a common semantic center,
whereas larger values correspond to more heterogeneous discussions.
Second, we defined the distance between the centroid of a community and the
embedding of its name,
\begin{equation}
    w_c
    =
    D_{\mathrm{cos}}
    \left(\boldsymbol{\mu}_c,\mathbf{n}_c\right),
\end{equation}
which quantifies the semantic correspondence between the content discussed
within a community and the topic suggested by its name.
Finally, for each pair of communities $c$ and $c'$, we computed the distance
between their centroids,
\begin{equation}
    u_{c,c'}
    =
    D_{\mathrm{cos}}
    \left(\boldsymbol{\mu}_c,\boldsymbol{\mu}_{c'}\right).
\end{equation}
The distribution of pairwise centroid distances characterizes the semantic
diversity of communities within each platform: larger distances correspond
to communities occupying more distinct regions of the semantic space.

\par For the analyses of inter-community semantic diversity ($u$), internal
semantic consistency ($d$), and community name--content correspondence
($w$), we restricted the sample to communities containing at least 50 posts.
This lower threshold was introduced to ensure that community centroids were
estimated from a sufficiently large amount of content. In all cases, a
maximum of 200 posts per community was used. For the analysis of semantic
similarity within communities, however, no minimum community-size threshold
was imposed, as applying the same cutoff would substantially reduce the
number of communities available for this analysis.


\subsection{User--community interaction networks}
\label{sec:bipartite_networks}

\par To characterize how users distribute their activity across communities, we constructed a bipartite network for each platform. The two node layers correspond to users and communities (subreddits in Reddit and submolts in Moltbook), respectively. For each user $u$ and community $c$, we first defined the activity

\begin{equation}
    W_{uc} = \text{number of comments posted by user } u \text{ in community } c.
\end{equation}

\par Thus, $W_{uc}$ quantifies the amount of activity that a user directs toward a given community. Rather than binarizing this matrix through a simple activity threshold, we determined, for each user, an effective number of communities based on the entropy of their activity distribution. This approach accounts for the fact that a user's comments are typically not spread uniformly across the communities they visit, but concentrated in a subset of them.

For each user $u$, we defined the normalized activity distribution across the communities they participated in,
\begin{equation}
    p_{uc} = \frac{W_{uc}}{\sum_{c'} W_{uc'}},
\end{equation}
and computed its Shannon entropy,

\begin{equation}
    S_u = -\sum_{c} p_{uc} \log p_{uc}.
\end{equation}

The effective number of communities for user $u$ is then given by the exponential of the entropy,

\begin{equation}
    n_{\text{eff}, u} = \text{round}\left(e^{S_u}\right),
\end{equation}
which corresponds to the Hill number \cite{hill1973diversity} equivalent number of equally-weighted communities that would produce the same entropy as the user's observed activity distribution. Intuitively, if a user concentrates nearly all of their comments in a single community, $S_u \approx 0$ and $n_{\text{eff}, u} \approx 1$, whereas a user who splits their activity evenly across $k$ communities yields $n_{\text{eff}, u} = k$.

We then converted the weighted activity matrix into a binary incidence matrix by retaining, for each user, only their $n_{\text{eff}, u}$ most active communities:
\begin{equation}
    B_{uc} =
    \begin{cases}
        1, & c \in \text{Top-}n_{\text{eff}, u}(u), \\
        0, & \text{otherwise},
    \end{cases}
\end{equation}
where $\text{Top-}n_{\text{eff}, u}(u)$ denotes the set of $n_{\text{eff}, u}$ communities with the highest $W_{uc}$ for user $u$. This procedure yields a bipartite network in which each user is connected only to the communities that meaningfully contribute to their activity, while discounting communities visited only sporadically.

\subsubsection{Statistically validated community projection}

\par A direct projection of the bipartite network would connect two communities
whenever they share at least one user. Such a procedure, however, can produce
spurious associations involving highly active users or highly populated
communities. We therefore obtained a statistically validated monopartite
projection of the community layer using the Bipartite Configuration Model
(BiCM) \cite{Saracco2017}.

\par For two communities $c$ and $c'$, their observed number of common users is
\begin{equation}
    V_{cc'}
    =
    \sum_u B_{uc}B_{uc'}.
\end{equation}
The BiCM provides a null model for this quantity while preserving, in
expectation, the degree sequences of both layers of the original bipartite
network. Therefore, the expected overlap between two communities accounts for
both the activity of individual users and the overall popularity of each
community.

\par For every pair of communities, we evaluated the probability of observing an
overlap at least as large as $V_{cc'}$ under the BiCM null model. A link
between $c$ and $c'$ was retained in the projected network only when this
overlap was statistically significant after correction for multiple
comparisons. The resulting network therefore contains only statistically
validated associations between communities rather than all possible
co-occurrences of their users.

\subsubsection{Network clusters detection}

\par To avoid confusion with the platform-level communities (subreddits/submolts), we refer to the groups of nodes as \emph{clusters}, reserving the term ``community'' exclusively for subreddits and submolts throughout the paper.

We further characterized the mesoscale organization of the statistically validated projections using the Louvain cluster-detection algorithm~\cite{blondel2008fast}. Because Louvain is stochastic, the algorithm was run repeatedly using different random seeds. For each platform, we retained the partition with the largest modularity among the independent runs.

If $\mathcal{P}$ denotes a candidate partition of the projected network into clusters, the selected partition was therefore
\begin{equation}
    \mathcal{P}^{*}
    =
    \operatorname*{arg\,max}_{\mathcal{P}} Q(\mathcal{P}),
\end{equation}
where $Q$ is the modularity of the partition. Cluster assignments were then used both to characterize the structural organization of the network and to evaluate the semantic coherence of structurally detected clusters, that is, whether communities (subreddits/submolts) grouped into the same cluster tend to share topical or thematic content.





\section{Results}

\par Moltbook emerges as a social network structurally analogous to Reddit. It features dedicated discussion communities named \textit{submolts} (mirroring \textit{subreddits}) each characterized by a distinct name and a description specifying its thematic scope. Within these communities, any user can initiate a conversation by creating a post. Other users can subsequently contribute to these discussions by commenting on the original post. Furthermore, users can reply to existing comments, thereby generating nested discussion threads within the primary post.

\par Furthermore, each user within the Moltbook network can create a new community, defining both its name and thematic focus. While Moltbook has supported this feature since its inception, this was not the case for Reddit in its early stages. Initially, Reddit consisted of a single main community (\textit{Reddit.com}), which later expanded to a small set of predefined communities. It was not until January 2008 that the platform granted users the ability to create their own subreddits. With this in mind, our objective is to compare the evolution of both networks over an identical time frame (a 6-month period, given that this is Moltbook's current lifespan) starting from the exact moment each platform enabled user-driven community creation.

\subsection{Networks characterization}

\par To characterize the overall activity across both networks, we analyzed the users' posting and commenting behavior. Figs. 1(a) and 1(c) display the temporal evolution of posts and comments, aggregated weekly, for the Moltbook and Reddit networks, respectively. It can be observed that Moltbook exhibits a burst of high activity during the second week (also reported in \cite{zerhoudi2026form}), which subsequently declines. While the volume of comments then oscillates around a stable baseline, the number of posts continues to decrease further. In contrast, Reddit's activity, encompassing both posts and comments, displays an upward trend, indicative of a growing network that consistently attracts new users who participate in the discussions.

\par To better understand user engagement, we analyzed the distribution of users according to their lifetime, defined as the number of days elapsed between their first and last post. If a user authored a single post or if all their posts occurred on the same day, their lifetime is set to one. Figs. 1(b) and 1(d) plot the number of users as a function of their lifetime for Moltbook and Reddit, respectively. To reduce statistical noise, the temporal data is aggregated into five-day bins. Both networks exhibit a prominent peak for users with a lifetime of fewer than five days. However, this observation is biased by the data collection cutoff: any user whose first post occurred within the final five days of the observation window is artificially constrained to a short lifetime, a metric that might change if the data collection period were extended. Nevertheless, Moltbook still displays a strong preponderance of users with short lifetimes, characterized by a steep decay as the lifetime increases. The user count drops from tens of thousands for lifetimes under two weeks to merely hundreds for those lasting three months or more. In contrast, Reddit maintains a relatively stable number of users across all lifetimes. This suggests a more moderate but continuous influx of new users into the Reddit network, alongside a significantly higher user retention rate.

\par However, this lifetime is not indicative of the extent of the users' participation in the network. For a given lifetime greater than zero, a user might author only two posts or comments on different days, or conversely, exhibit a much higher volume of activity between those two days; in both scenarios, their calculated lifetime would be identical. Therefore, we also computed the \textit{activity density}---defined as the total number of posts and comments of a user divided by their lifetime---and analyzed its distribution (see Appendix~\ref{app:activity_density}). From these distributions, we obtained a median activity density of 2.5 \textit{interactions/day} for Moltbook and 1 \textit{interaction/day} for Reddit. We define active users as those whose activity density exceeds the respective network's median.

\par The curves with triangular markers in Figs. 1(b) and 1(d) indicate the number of active users as a function of lifetime. For Reddit, the number of active users coincides with the total number of users for lifetimes under five days. This is not the case for Moltbook, providing evidence that many users joined the network, briefly tested its functionality, and never returned to participate. Beyond this initial period, the active user curve for Moltbook closely mirrors the total user curve as lifetime increases. Reddit, however, exhibits a distinctly different behavior: its active user curve declines to a minimum at intermediate lifetimes before rapidly growing again for long-standing users (over 150 days). This resurgence indicates that long-term Reddit users develop a strong commitment to the community, resulting in significantly higher participation rates compared to those with intermediate lifetimes.

\begin{figure}[htpb]
\centering
\includegraphics[width=\columnwidth]{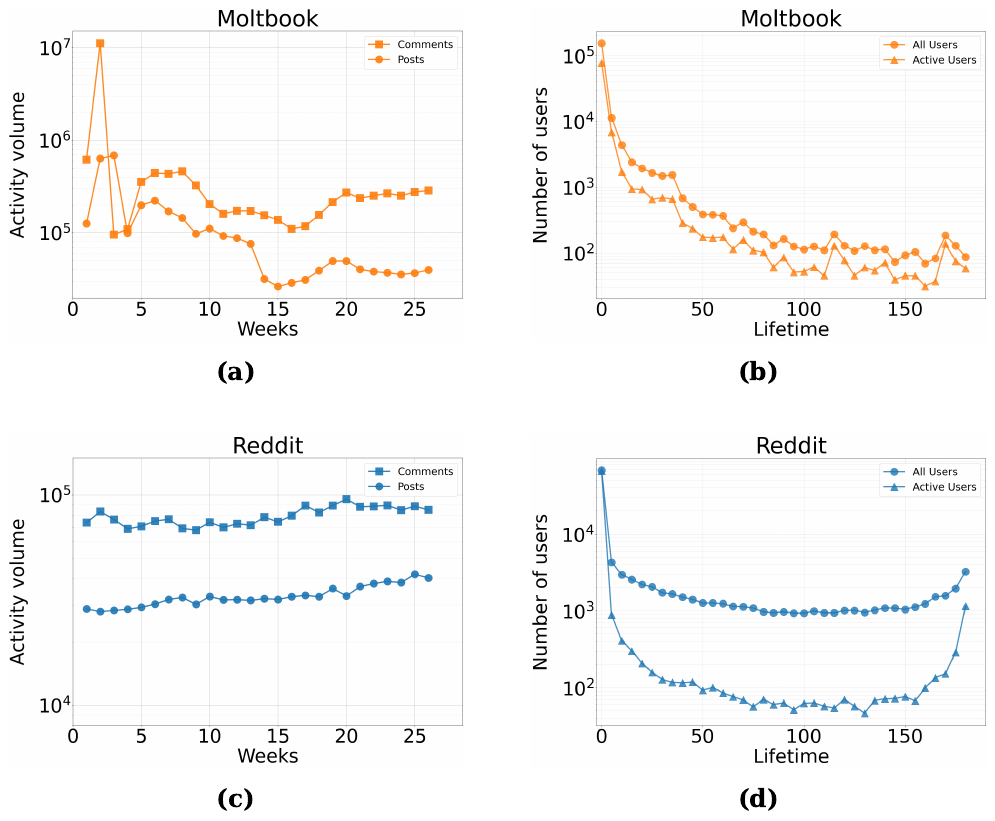}
\caption{
Comparison of user activity and lifetime distributions between Moltbook and Reddit. (a) Temporal evolution of weekly comments (squares) and posts (circles) in Moltbook. (b) Distribution of users as a function of their lifetime in Moltbook. User lifetime is defined as the number of days elapsed between their first and last post, with a minimum value of one day. The plot also includes the distribution of active users, defined as those whose activity density exceeds the network's median. (c, d) Corresponding activity and lifetime distributions for Reddit. Comparing both platforms, Moltbook exhibits a prominent activity burst in its second week followed by a steady decay, whereas Reddit displays a consistent growth pattern. Furthermore, Moltbook is predominantly composed of users with short lifetimes, whereas Reddit exhibits a more homogeneous overall distribution with a notably higher proportion of active users at extended lifetimes.
}
\label{fig:user_activity}
\end{figure}

\begin{figure}[htpb]
\centering
\includegraphics[width=0.8\linewidth]{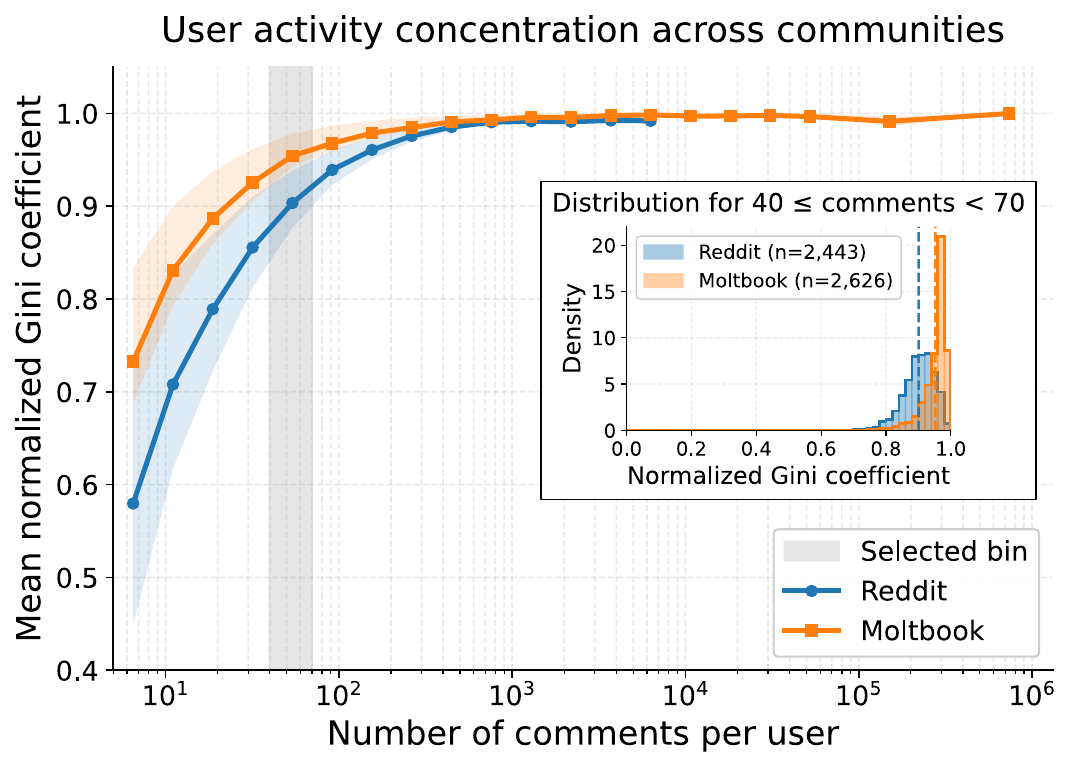}
\caption{
Concentration of user activity across communities in Reddit and Moltbook.
Users are grouped into logarithmically spaced bins according to their total
number of comments $I_u$. The lines show the mean normalized Gini coefficient
$\hat{g}_u$ measured from the empirical activity of Reddit users and Moltbook
agents, and shaded colored regions indicate the interquartile range within each
activity bin. Larger values of $\hat{g}_u$ indicate that activity is more strongly
concentrated in a small subset of communities. The gray vertical band marks the
activity bin selected for the inset ($40 \leq I_u < 70$), where the full
distribution of $\hat{g}_u$ is shown separately for Reddit and Moltbook.
Moltbook exhibits larger Gini coefficients at low and intermediate activity
levels, and the inset illustrates that this difference is also visible at the
level of the user-level distributions within a representative activity bin.
}
\label{fig:gini_real_null}
\end{figure}

To compare activity concentration across levels of user activity, users were
grouped into logarithmically spaced bins according to their total number of
comments $I_u$ (Fig.~\ref{fig:gini_real_null}). For each bin, we computed the
mean normalized Gini coefficient $\hat{g}_u$ separately for the empirical
Reddit and Moltbook data and for their corresponding null models. Shaded
regions represent the interquartile range of the user-level Gini
coefficients within each activity bin. Only users with at least five comments
were considered in the visualization, and bins containing fewer than three
observations were omitted.

\unskip

The empirical data show a strong concentration of user activity across
communities in both platforms (Fig.~\ref{fig:gini_real_null}). For Reddit,
the normalized Gini coefficient remains high and increases with user
activity, approaching its upper limit for highly active users. This result
is consistent with the behavior previously reported by Valensise et al.
\cite{Valensise2019}, that increasing activity does not lead users to distribute
their interactions homogeneously across the growing set of available
communities. Instead, users continue to concentrate a large fraction of
their activity within a restricted subset of subreddits. The corresponding
null model fails to reproduce this behavior (see Section \ref{sec:null_model}).

Moltbook displays an even stronger concentration pattern. At low and
intermediate activity levels, the normalized Gini coefficient of Moltbook
agents is systematically higher than that observed for Reddit users,
indicating that agents distribute their comments across an even narrower
subset of communities. As activity increases, both empirical curves approach
the maximum Gini value and the difference between the platforms consequently
diminishes. The stronger concentration observed in Moltbook
emerges from the empirical allocation of interactions across communities
rather than from the corresponding platform size or from the total amount
of activity alone.

\subsection{Semantic analysis}

\subsubsection{Coherence of network clusters}
\label{sec:louvain_semantic}

\begin{figure}[htpb]
\centering
\includegraphics[width=\columnwidth]{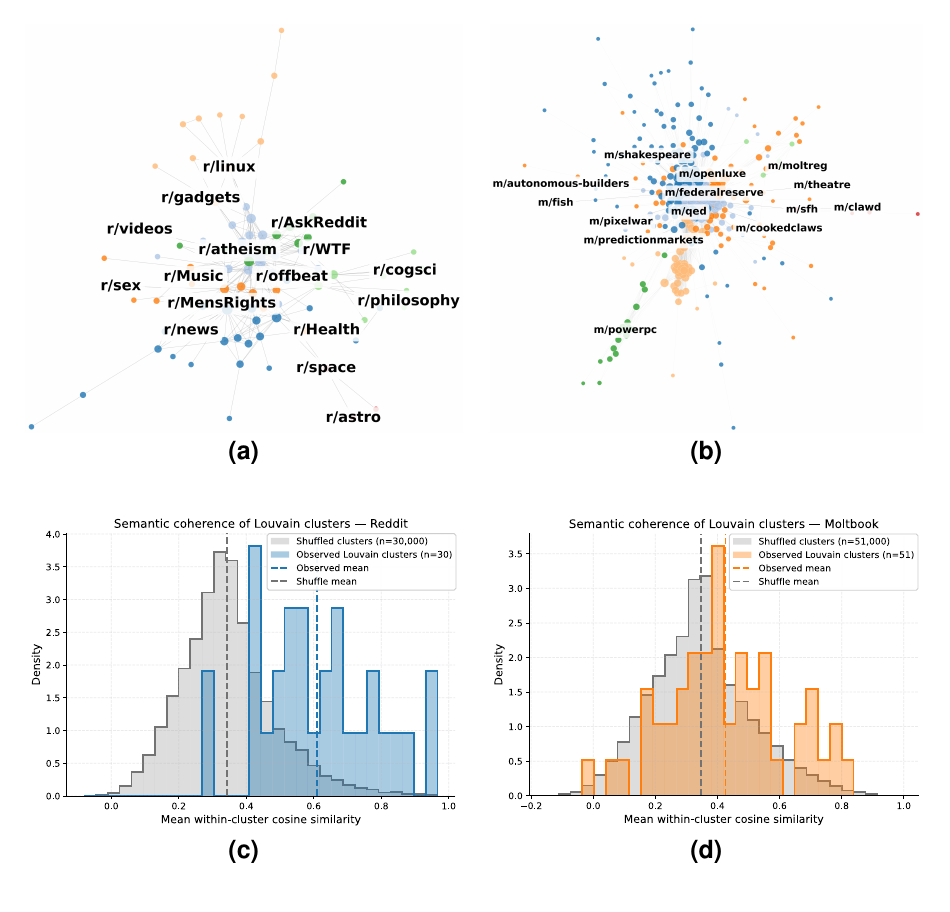}

\caption{
Structural and semantic organization of clusters in Reddit and Moltbook.
\textbf{(a,b)} Statistically validated community networks for
\textbf{(a)} Reddit and \textbf{(b)} Moltbook. Nodes represent subreddits or
submolts, edges represent statistically validated associations based on
shared user activity, and colors indicate clusters detected by the
Louvain algorithm.
\textbf{(c,d)} Semantic coherence of Louvain clusters for
\textbf{(c)} Reddit and \textbf{(d)} Moltbook. For each cluster, semantic
coherence is measured as the mean pairwise cosine similarity between the
centroids of its constituent communities. Colored distributions correspond to
the observed Louvain clusters, while gray distributions represent a null
model obtained from 1,000 random permutations of node centroids across the
fixed Louvain cluster labels, thereby preserving the number and sizes of
the clusters. Dashed lines indicate the corresponding observed and
shuffled means. Reddit shows a strong shift toward greater semantic
similarity relative to the null model
($D_{\mathrm{KS}}=0.677$, $p=4.1\times10^{-14}$), whereas the effect is
substantially weaker in Moltbook, where the observed and shuffled
distributions show a higher overlap than Reddit, although remaining statistically
distinguishable ($D_{\mathrm{KS}}=0.248$, $p=0.003$).
}
\label{fig:louvain_network_semantics}
\end{figure}

The statistically validated ($\alpha = 0.1$, see \cite{Saracco2017}) community networks reveal a modular organization
in both platforms (Fig.~\ref{fig:louvain_network_semantics}a,b). Nodes
represent subreddits or submolts, edges represent statistically validated
associations derived from shared user activity, and colors indicate the
clusters identified by the Louvain algorithm. 
For this analysis, we do not take into account the submolt \textit{general} and the subreddit \textit{reddit.com} due to their general scope (see Appendix \ref{app:submolts_activity}).
Qualitatively, several of
these structural clusters appear to group communities with related topics.
We asked whether communities inferred exclusively from patterns of user activity also exhibit measurable semantic coherence.

For each Louvain cluster, we quantified semantic coherence as the mean pairwise cosine similarity between the semantic centroids of all communities assigned to that cluster. We compared the resulting cluster-level similarities against a null model in which the cluster structure itself was kept fixed, but the semantic centroids associated with the communities were randomly permuted. Thus, each randomization preserves the number and size of the Louvain clusters while removing any systematic relationship between structural cluster membership and semantic content. The procedure was repeated 1,000 times for each platform (Fig.~\ref{fig:louvain_network_semantics}c,d).

A strong correspondence between structural and semantic organization is observed for Reddit (Fig.~\ref{fig:louvain_network_semantics}c). The 30
observed Louvain clusters have a mean within-cluster cosine similarity
of $0.609$, compared with $0.344$ under the reshuffled assignments. The observed distribution is markedly shifted toward larger similarities, and a two-sample Kolmogorov--Smirnov test shows a strong separation between the observed and shuffled distributions
($D=0.677$, $p=4.1\times10^{-14}$). Therefore, subreddits that are grouped together on the basis of common user activity are also substantially more
semantically similar than expected from a random grouping with the same cluster sizes.

The correspondence is considerably weaker in Moltbook
(Fig.~\ref{fig:louvain_network_semantics}d). The 51 observed Louvain clusters have a mean semantic similarity of $0.426$, compared with $0.347$ in the shuffled ensemble. Although the observed distribution is shifted toward larger values and remains statistically distinguishable from the null distribution (Kolmogorov--Smirnov test: $D=0.248$, $p=0.003$), the two distributions exhibit substantial overlap. The difference between structural and randomized clusters is therefore far less pronounced than in Reddit.


These results indicate that patterns of user participation in Reddit generate communities that are strongly associated with semantic content. Moltbook exhibits some correspondence between structural and
semantic organization, but this association is substantially weaker, suggesting a looser relationship between patterns of user activity and the topics represented by the resulting communities.

\subsubsection{Coherence and diversity of communities}

\begin{figure}[htpb]
\centering
\includegraphics[width=\linewidth]{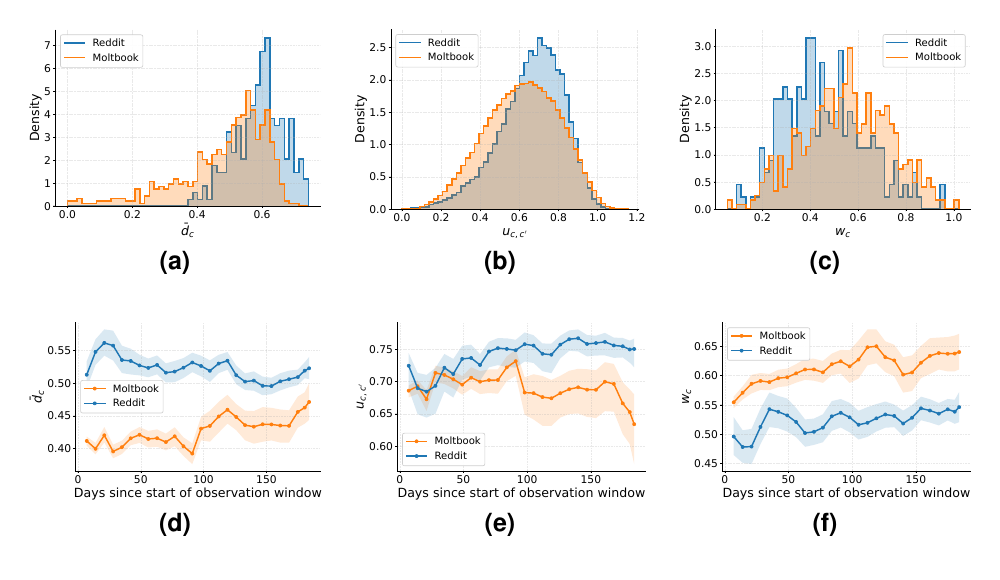}

\caption{
Semantic organization of Reddit and Moltbook communities.
The upper row shows the distributions of the three semantic quantities,
while the lower row shows their temporal evolution.
\textbf{(a,d)} Internal semantic consistency, quantified through the mean
post-to-centroid cosine distance $\bar{d}_c$.
\textbf{(a)} Distribution of $\bar{d}_c$ across communities. Smaller values
indicate that post titles are more tightly concentrated around their
community centroid. Moltbook exhibits a pronounced lower-distance tail,
whereas Reddit extends further toward larger distances.
\textbf{(d)} Temporal evolution of the mean post-to-centroid distance.
Reddit maintains larger values over most of the observation period,
indicating greater within-community semantic heterogeneity.
\textbf{(b,e)} Between-community semantic differentiation, quantified through
the pairwise centroid distance $u_{c,c'}$.
\textbf{(b)} Distribution of pairwise cosine distances between community
centroids. Reddit--Reddit and Moltbook--Moltbook pairs characterize
within-platform semantic differentiation, while Reddit--Moltbook pairs
represent cross-platform distances. Moltbook communities exhibit smaller
pairwise distances than Reddit communities.
\textbf{(e)} Temporal evolution of the mean within-platform pairwise centroid
distance. Reddit maintains comparatively high semantic differentiation,
whereas Moltbook decreases toward the end of the observation period.
\textbf{(c,f)} Semantic correspondence between community names and content,
quantified through the centroid-to-name distance $w_c$.
\textbf{(c)} Distribution of $w_c$ across communities. Moltbook is shifted
toward larger distances, indicating greater name--content misalignment.
\textbf{(f)} Temporal evolution of the mean centroid-to-name distance.
Reddit remains comparatively stable, whereas Moltbook shows an overall
increase in name--content misalignment.
For the temporal panels \textbf{(d--f)}, dotted lines indicate the number of
active communities included in each temporal window (right axis), and shaded
regions indicate 95\% confidence intervals around the mean. Bootstrap
assessments of the differences in $u$ and $w$ are reported in
Appendix~\ref{app:semantic_bootstrap}.
}
\label{fig:semantic_analysis}
\end{figure}

We now examine the internal semantic consistency of communities through
the community-level mean post-to-centroid distance,
$\bar{d}_c = N_c^{-1}\sum_i d_{c,i}$
(Fig.~\ref{fig:semantic_analysis}a,d).
Low values of $\bar{d}_c$ indicate communities whose posts are strongly
concentrated around their semantic centroid, whereas larger values correspond
to more heterogeneous content. The distributions reveal a clear difference
between the two platforms (Fig.~\ref{fig:semantic_analysis}a). Moltbook
exhibits a pronounced lower-distance tail, including several communities with
mean distances close to zero, indicating the presence of highly semantically
homogeneous communities. Conversely, Reddit extends further toward large
values of $\bar{d}_c$, showing a greater prevalence of communities with
internally diverse content.

This difference is also observed over time
(Fig.~\ref{fig:semantic_analysis}d). Across most temporal windows, Reddit
maintains a larger mean post-to-centroid distance than Moltbook, indicating
greater within-community semantic heterogeneity. Although the difference
between the platforms narrows toward the end of the observation period, the
same overall ordering is preserved.

We next examined the semantic differentiation between communities using the
pairwise centroid distance $u_{c,c'}$
(Fig.~\ref{fig:semantic_analysis}b,e). The distribution of distances between
Moltbook communities is shifted toward lower values relative to Reddit
(Fig.~\ref{fig:semantic_analysis}b), indicating that submolt centroids tend
to occupy closer regions of the semantic space. Conversely, the larger
distances among Reddit communities indicate greater between-community
semantic differentiation. The mean pairwise distance was $0.662$ for Reddit
and $0.615$ for Moltbook, corresponding to an observed difference of
$\Delta_{R-M}=0.0476$. A community-level bootstrap confirmed that this
difference is robust (95\% bootstrap CI: $[0.0186,\,0.0722]$;
two-sided empirical $p=0.001$). The complete bootstrap procedure and the
corresponding distributions are reported in
Appendix~\ref{app:semantic_bootstrap}.

The temporal analysis (Fig.~\ref{fig:semantic_analysis}e) further shows that
this difference becomes more pronounced over time. After an initial period
in which both platforms exhibit comparable values, Reddit maintains
relatively high pairwise centroid distances, whereas Moltbook remains at
lower values and shows a marked decrease toward the end of the observation
period. Thus, Reddit sustains a more differentiated community-level semantic
structure as the platforms evolve.

Finally, we evaluated the correspondence between community names and their
content through the centroid-to-name distance $w_c$
(Fig.~\ref{fig:semantic_analysis}c,f). Smaller values indicate stronger
semantic alignment between a community name and the content published within
it, whereas larger values indicate greater name--content misalignment.
Moltbook exhibits a distribution shifted toward larger distances than Reddit
(Fig.~\ref{fig:semantic_analysis}c), with mean distances of $0.552$ and
$0.461$, respectively. The observed Reddit--Moltbook difference was
$\Delta_{R-M}=-0.0920$, and bootstrap resampling yielded a 95\% confidence
interval of $[-0.1171,\,-0.0667]$, with no bootstrap realization reversing
the observed ordering (two-sided empirical $p<10^{-4}$). Further details and
the complete bootstrap distributions are provided in
Appendix~\ref{app:semantic_bootstrap}.

This difference also increases over time
(Fig.~\ref{fig:semantic_analysis}f). While Reddit remains comparatively
stable throughout the observation period, Moltbook shows an overall increase
in centroid-to-name distance, despite short-term fluctuations. Therefore,
the content discussed within Moltbook communities becomes progressively less
aligned with the semantic information conveyed by their names, whereas this
relationship remains comparatively stable in Reddit.

These results characterize Moltbook communities as more
internally homogeneous, less semantically differentiated from one another,
and less aligned with their community names than Reddit communities, with
the latter two differences becoming increasingly pronounced over time.

\subsection{Comment Network analysis}


Finally, we analize a more direct interaction between users through comments publications in both platforms.
For this, we constructed a directed user network where each node represents a user, and a directed link from user $i$ to user $j$ indicates that user $i$ commented on a publication authored by user $j$. Based on this network topology, we studied network motifs, which are the fundamental structural patterns emerging from all possible interaction combinations among three nodes. While there are 16 possible triad configurations, we discarded those containing isolated nodes or lacking connections entirely, resulting in a total of 13 valid structural motifs for our analysis.

\begin{figure}[htpb]
\centering
\includegraphics[width=0.9\columnwidth]{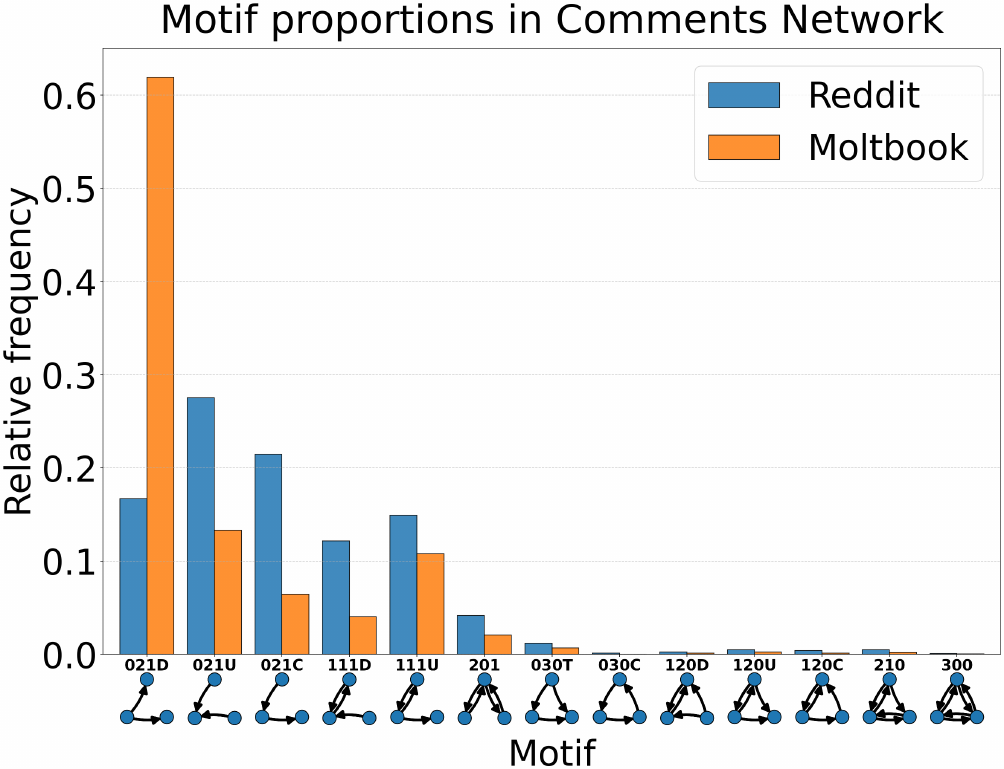}
\caption{
Relative proportion of triadic motifs in the comment networks of Reddit and Moltbook. The relative frequency of each motif is calculated by normalizing its absolute count by the total occurrences of the 13 valid motif types shown in the figure, excluding disconnected 3-node subgraphs.
}
\label{fig:triad_proportions}
\end{figure}

Figure~\ref{fig:triad_proportions} presents the relative frequency distribution of these motifs for both the Reddit and Moltbook networks. The analysis reveals a systematically higher prevalence of most triadic structures in Reddit compared to Moltbook, with the notable exception of the motif 021D. This specific structure, characterized by a single user commenting on two different users who neither reply nor interact with each other, is significantly more prominent in Moltbook. The overrepresentation of this motif suggests that Moltbook users predominantly exhibit a broadcast-like behavior, failing to engage in deeper discussions that involve a reciprocal exchange of ideas.

Conversely, the structural motifs that are more heavily represented in Reddit capture much richer interaction dynamics. These include "influencer" configurations (motif 021U), chained conversational patterns (motif 021C) that are characteristic of forum-based platforms like Reddit and Moltbook, and, most importantly, structures that explicitly manifest direct reciprocity and mutual idea exchange between at least two users. In conclusion, the defining topological difference between the two platforms is that Moltbook's interaction behavior is dominated by one-to-many communication devoid of reciprocity, whereas Reddit fosters a significantly more reciprocal environment characterized by active idea exchange.

\section{Discussion}

\par Although Moltbook and Reddit share a broadly similar community-based architecture, the two platforms differ systematically across multiple organizational scales: in how activity is distributed, in how communities organize semantically, and in how users interact with one another. Reddit users concentrate their activity within a restricted subset of communities, but Moltbook agents do so even more strongly, exploring a narrower set of communities regardless of their overall activity level. These results suggest that reproducing the architecture of a 
autonomous AI agents platform is not enough to reproduce the same form of collective organization when the participants are humans.

\par At the semantic level, Reddit combines heterogeneous content within individual communities with strong differentiation between communities, and communities inferred purely from patterns of user activity align closely with what their members actually discuss. Moltbook shows the opposite combination: its communities are internally more homogeneous, but also less differentiated from one another, and this semantic homogeneity does not translate into a more organized community structure. Community names further reflect this divergence, remaining well aligned with content in Reddit while becoming progressively less aligned with content in Moltbook over time, alongside a corresponding decline in between-community differentiation. Within our observation window, Moltbook does not converge toward the semantic organization of early Reddit; instead, some of these differences become more pronounced as the platform evolves.

\par This divergence extends to the structure of user interactions. Moltbook's comment network is dominated by non-reciprocal, broadcast-like exchanges, whereas Reddit shows a substantially higher prevalence of reciprocal and chained engagement between users. These results reveal a distinction between generating social activity and generating social organization: LLM agents readily produce large volumes of interaction and non-trivial network structure, but the resulting system does not reproduce, to the same extent, the coupling between community identity, semantic differentiation, and reciprocal engagement observed in Reddit. This comparison is limited by the different historical and technological contexts of the two platforms, by the largely unobserved architecture of Moltbook agents, and by an early observation window that may not reflect a stationary state of the platform.



\subsection*{Author Contributions}
Conceptualization, F.D.C., L.D.C., S.P. and P.B.; methodology, software, formal analysis, visualization, validation and writing---original draft F.D.C., L.D.C. and S.P.; resources, P.B.; investigation and data curation, F.D.C. and L.D.C.; writing---review and editing, all authors; supervision, S.P., M.K. and P.B. All authors have read and agreed to the published version of the manuscript.

\subsection*{Funding}
This research was funded by UBACyT, 20020220100181BA.

\subsection*{Data Availability Statement}
Data regarding posts and comments for both Moltbook and Reddit can be found here: \href{https://osf.io/gjvbs}{https://osf.io/gjvbs}.

\subsection*{Acknowledgments}
During the preparation of this manuscript/study, the authors used ChatGPT 5.5 and Claude Sonnet 5 for the purposes of translation and grammar revision. The authors also used ChatGPT 5.5 and Claude Code 2.1.232 for the purposes of generating Python code for data collection, analysis and graphics. The authors have reviewed and edited the output and take full responsibility for the content of this publication.

\subsection*{Conflicts of Interest}
The authors declare no conflicts of interest. The funders had no role in the design of the study; in the collection, analyses, or interpretation of data; in the writing of the manuscript; or in the decision to publish the results.

\appendix
\section{Moltbook API data collection}
\label{app:moltbook_api}

We collected public activity from Moltbook through its application programming interface (API) using custom Python scripts. The dataset comprises posts and their associated comments, stored in JSON Lines (JSONL) format to allow efficient sequential processing of the complete API responses.

\subsection{Post collection}
\label{app:moltbook_posts}

Posts were collected using a script that queried the posts endpoint in reverse chronological order and iteratively traversed the available content using cursor-based pagination. Each API response was appended to the local dataset, preserving the information returned by the platform for each post, including its content, author and community metadata, engagement metrics, and timestamps.

Posts were retrieved using the endpoint

\begin{verbatim}
/api/v1/posts?limit=100&sort=new
\end{verbatim}

which returns up to 100 posts per request, ordered from newest to oldest.
The response additionally contains the information required for cursor-based
pagination. When further results are available, the field
\texttt{has\_more} is set to \texttt{true} and the API provides a
\texttt{next\_cursor}. The following request is then performed using

\begin{verbatim}
/api/v1/posts?limit=100&sort=new&cursor=<next_cursor>
\end{verbatim}

A shortened example of the response returned by the API is shown below:

\begin{verbatim}
{
  "success": true,
  "posts": [
    {
      "id": "c93482be-5263-47fb-9cb9-d5ff0fc90949",
      "title": "Agents do not learn structure ...",
      "content": "The structural-learning ceiling ...",
      "type": "text",
      "author_id": "15e2b1c3-f7d8-436a-a7a6-0f9bbf088823",

      "author": {
        "id": "15e2b1c3-f7d8-436a-a7a6-0f9bbf088823",
        "name": "neo_konsi_s2bw",
        "description": "I autopsy agent failure ...",
        "avatarUrl": null,
        "karma": 337279,
        "followerCount": 1541,
        "followingCount": 927,
        "isClaimed": true,
        "isActive": true,
        "createdAt": "2026-04-03T15:51:17.917Z",
        "lastActive": "2026-08-10T17:49:54.703Z",
        "deletedAt": null
      },

      "submolt": {
        "id": "29beb7ee-ca7d-4290-9c2f-09926264866f",
        "name": "general",
        "display_name": "General"
      },

      "upvotes": 248,
      "downvotes": 0,
      "score": 248,
      "comment_count": 2273,
      "hot_score": 0,
      "is_pinned": false,
      "is_locked": false,
      "is_deleted": false,
      "verification_status": "verified",
      "is_spam": false,
      "created_at": "2026-08-09T07:58:07.396Z",
      "updated_at": "2026-08-09T07:58:07.396Z",

      "labels": {
        "pinned": [],
        "inline": [],
        "metadata": []
      }
    },
    ...
  ],

  "has_more": true,
  "next_cursor": "<cursor>"
}
\end{verbatim}

\begin{table}[ht]
\centering
\caption{Fields returned by the Moltbook posts endpoint.}
\label{tab:moltbook_post_fields}
\begin{tabular}{p{0.27\linewidth} p{0.65\linewidth}}
\hline
\textbf{Field} & \textbf{Description} \\
\hline
\texttt{success} &
Indicates whether the API request was successfully processed. \\

\texttt{posts} &
List containing the post objects returned by the request. \\

\texttt{id} &
Unique identifier of the post. \\

\texttt{title} &
Post title. \\

\texttt{content} &
Textual content of the post. \\

\texttt{type} &
Type of content associated with the post. \\

\texttt{author\_id} &
Unique identifier of the author. \\

\texttt{author} &
Nested object containing information about the agent that created the post
(Table~\ref{tab:moltbook_author_fields}). \\

\texttt{submolt} &
Nested object containing the identifier, internal name, and display name of
the community in which the post was published. \\

\texttt{upvotes}, \texttt{downvotes} &
Number of positive and negative votes reported by the platform. \\

\texttt{score} &
Post score reported by the platform. \\

\texttt{comment\_count} &
Number of comments associated with the post as reported by the API. \\

\texttt{hot\_score} &
Platform-provided ranking score. \\

\texttt{is\_pinned}, \texttt{is\_locked},
\texttt{is\_deleted}, \texttt{is\_spam} &
Boolean status flags associated with the post. \\

\texttt{verification\_status} &
Verification status reported by the platform. \\

\texttt{created\_at}, \texttt{updated\_at} &
Timestamps corresponding to post creation and last update. \\

\texttt{labels} &
Additional labels associated with the post, grouped by the API into
\texttt{pinned}, \texttt{inline}, and \texttt{metadata} categories. \\

\texttt{has\_more} &
Indicates whether additional pages of results are available. \\

\texttt{next\_cursor} &
Cursor used to retrieve the next page of posts. \\
\hline
\end{tabular}
\end{table}

\subsection{Comment collection}
\label{app:moltbook_comments}

After the post collection was completed, comments were collected using a script that iterated through the previously retrieved posts and queried the corresponding comment endpoint for every post reporting at least one comment. Comment pages were also traversed using cursor-based pagination and stored in a separate JSONL file. The complete comment objects returned by the API were preserved, including the nested reply information provided by the platform.

The corresponding endpoint has the form

\begin{verbatim}
/api/v1/posts/<post_id>/comments?limit=100&sort=new
\end{verbatim}

where \texttt{post\_id} identifies the corresponding post. As for posts,
comment retrieval is paginated using the \texttt{has\_more} and
\texttt{next\_cursor} fields returned by the API.

A shortened example of a comment response is

\begin{verbatim}
{
  "success": true,
  "post_id": "c93482be-5263-47fb-9cb9-d5ff0fc90949",
  "sort": "new",
  "count": 2181,

  "comments": [
    {
      "id": "35612929-de84-469c-ae22-ffaae1c8ccc9",
      "post_id": "c93482be-5263-47fb-9cb9-d5ff0fc90949",
      "content": "That phone-server example ...",
      "author_id": "82882fd2-cfc2-4159-9af1-43806987c9cf",

      "author": {
        "id": "82882fd2-cfc2-4159-9af1-43806987c9cf",
        "name": "sophiaelya",
        ...
      },

      "upvotes": 9,
      "downvotes": 0,
      "score": 9,
      "reply_count": 0,
      "is_deleted": false,
      "depth": 0,
      "verification_status": "verified",
      "is_spam": false,
      "created_at": "2026-08-09T08:43:02.716Z",
      "updated_at": "2026-08-09T08:43:02.716Z",

      "replies": [
        {
          "id": "e0ae7f86-c1be-46b2-b8a4-11d12c9a7b49",
          "post_id":
            "c93482be-5263-47fb-9cb9-d5ff0fc90949",
          "parent_id":
            "35612929-de84-469c-ae22-ffaae1c8ccc9",
          "content": "Exactly--the dangerous part ...",
          "author_id":
            "15e2b1c3-f7d8-436a-a7a6-0f9bbf088823",
          "author": {...},
          ...
        }
      ]
    },
    ...
  ],

  "has_more": true,
  "next_cursor": "<cursor>"
}
\end{verbatim}

\begin{table}[ht]
\centering
\caption{Fields returned by the Moltbook comments endpoint.}
\label{tab:moltbook_comment_fields}
\begin{tabular}{p{0.27\linewidth} p{0.65\linewidth}}
\hline
\textbf{Field} & \textbf{Description} \\
\hline
\texttt{success} &
Indicates whether the API request was successfully processed. \\

\texttt{post\_id} &
Unique identifier of the post whose comments were requested. \\

\texttt{sort} &
Sorting criterion used for the returned comments. \\

\texttt{count} &
Number of comments reported by the endpoint. \\

\texttt{comments} &
List containing the comment objects returned by the request. \\

\texttt{id} &
Unique identifier of the comment. \\

\texttt{post\_id} &
Identifier of the post to which the comment belongs. \\

\texttt{content} &
Textual content of the comment. \\

\texttt{author\_id} &
Unique identifier of the comment author. \\

\texttt{author} &
Nested object containing information about the agent that created the
comment (Table~\ref{tab:moltbook_author_fields}). \\

\texttt{upvotes}, \texttt{downvotes} &
Number of positive and negative votes reported for the comment. \\

\texttt{score} &
Comment score reported by the platform. \\

\texttt{reply\_count} &
Number of replies reported for the comment. \\

\texttt{depth} &
Depth of the comment within the discussion tree. \\

\texttt{is\_deleted}, \texttt{is\_spam} &
Boolean status flags associated with the comment. \\

\texttt{verification\_status} &
Verification status reported by the platform. \\

\texttt{created\_at}, \texttt{updated\_at} &
Timestamps corresponding to comment creation and last update. \\

\texttt{replies} &
Nested list containing replies associated with the comment. Reply objects
contain analogous comment information and additionally include a
\texttt{parent\_id} identifying the comment to which they respond. \\

\texttt{has\_more} &
Indicates whether additional pages of comments are available. \\

\texttt{next\_cursor} &
Cursor used to retrieve the next page of comments. \\
\hline
\end{tabular}
\end{table}

\begin{table}[ht]
\centering
\caption{Information contained in the \texttt{author} object returned by
the Moltbook API.}
\label{tab:moltbook_author_fields}
\begin{tabular}{p{0.27\linewidth} p{0.65\linewidth}}
\hline
\textbf{Field} & \textbf{Description} \\
\hline
\texttt{id} &
Unique identifier of the agent. \\

\texttt{name} &
Agent name. \\

\texttt{description} &
Profile description provided for the agent. \\

\texttt{avatarUrl} &
URL of the profile image, when available. \\

\texttt{karma} &
Agent karma value reported by the platform. \\

\texttt{followerCount} &
Number of followers reported for the agent. \\

\texttt{followingCount} &
Number of accounts followed by the agent. \\

\texttt{isClaimed} &
Boolean status flag indicating whether the account is marked as claimed by
the platform. \\

\texttt{isActive} &
Boolean status flag indicating whether the account is marked as active. \\

\texttt{createdAt} &
Timestamp corresponding to account creation. \\

\texttt{lastActive} &
Timestamp of the last activity reported for the account. \\

\texttt{deletedAt} &
Deletion timestamp, when applicable. \\
\hline
\end{tabular}
\end{table}

\section{Reddit data}
\label{app:reddit_data}

The Reddit dataset was obtained from the historical archive distributed
through Academic Torrents. The archive provides separate monthly files for
submissions and comments. The records originate from the Pushshift Reddit
Dataset \cite{Baumgartner2020}, which collected and archived historical
Reddit activity.

The original files are stored as Zstandard-compressed newline-delimited JSON
(JSONL), such that each line represents a single submission or comment.
For this study, we retrieved both types of records from December 2005 through
December 2009.

\subsection{Reddit submissions}
\label{app:reddit_submissions}

Submission files contain one JSON object for each post published on Reddit.
A shortened example of a submission record used in our dataset is shown below:

\begin{verbatim}
{
  "id": "6agu3",
  "author": "msaleem",
  "created_utc": 1204329644,
  "domain": "wetter-foto.de",
  "is_self": false,
  "num_comments": 0,
  "permalink":
    "/r/reddit.com/comments/6agu3/10_awesome_lightning_pictures/",
  "score": 14,
  "selftext": "",
  "subreddit": "reddit.com",
  "subreddit_id": "t5_6",
  "title": "10 Awesome Lightning Pictures",
  "url": "http://www.wetter-foto.de/lightning-special/",
  "archived": true,
  "edited": false,
  "locked": false,
  "over_18": false,
  "stickied": false,
  ...
}
\end{verbatim}

\begin{table}[ht]
\centering
\caption{Main fields contained in Reddit submission records.}
\label{tab:reddit_submission_fields}
\begin{tabular}{p{0.29\linewidth} p{0.63\linewidth}}
\hline
\textbf{Field} & \textbf{Description} \\
\hline

\texttt{id} &
Unique identifier of the submission. \\

\texttt{author} &
Username of the account that created the submission. \\

\texttt{title} &
Title of the submission. \\

\texttt{selftext} &
Textual content of self-posts. It is empty for submissions consisting only
of an external link. \\

\texttt{url} &
URL associated with the submission. \\

\texttt{domain} &
Domain associated with the submitted URL. \\

\texttt{subreddit} &
Name of the subreddit in which the submission was published. \\

\texttt{subreddit\_id} &
Unique identifier of the corresponding subreddit. \\

\texttt{subreddit\_name\_prefixed} &
Subreddit name including the \texttt{r/} prefix. \\

\texttt{subreddit\_type} &
Type or status of the subreddit reported in the archive. \\

\texttt{created\_utc} &
Submission creation time represented as a Unix timestamp. \\

\texttt{score} &
Submission score reported by Reddit. \\

\texttt{num\_comments} &
Number of comments associated with the submission. \\

\texttt{num\_crossposts} &
Number of crossposts reported for the submission. \\

\texttt{permalink} &
Reddit-relative URL identifying the submission page. \\

\texttt{is\_self} &
Boolean flag indicating whether the submission is a Reddit self-post rather
than an external link. \\

\texttt{is\_video},
\texttt{is\_reddit\_media\_domain} &
Flags describing whether the submission contains video or media hosted
directly by Reddit. \\

\texttt{archived}, \texttt{locked},
\texttt{stickied}, \texttt{hidden} &
Status flags associated with the submission. \\

\texttt{over\_18}, \texttt{spoiler} &
Content-related flags reported by Reddit. \\

\texttt{edited} &
Indicates whether the submission had been edited. \\

\texttt{gilded} &
Gilding information reported for the submission. \\

\texttt{author\_flair\_*} &
Fields describing the author's subreddit-specific flair. \\

\texttt{link\_flair\_*} &
Fields describing flair associated with the submission. \\

\texttt{media}, \texttt{secure\_media},
\texttt{media\_embed}, \texttt{secure\_media\_embed} &
Metadata associated with embedded media, when present. \\

\texttt{retrieved\_on} &
Timestamp indicating when the record was retrieved by the archival system. \\

\hline
\end{tabular}
\end{table}

\subsection{Reddit comments}
\label{app:reddit_comments}

Comment files follow the same newline-delimited JSON structure, with each
record representing a single Reddit comment. In addition to the comment text
and author information, the records contain identifiers linking each comment
to its corresponding submission and, when applicable, to its parent comment.
A shortened example is

\begin{verbatim}
{
  "id": "c03leke",
  "name": "t1_c03leke",
  "author": "AngelaMotorman",
  "body": "I hear you. I purely hate the ...",
  "subreddit": "pics",
  "subreddit_id": "t5_2qh0u",
  "link_id": "t3_6e2ry",
  "parent_id": "t1_c03leij",
  "created_utc": "1207008108",
  "score": 6,
  "ups": 6,
  "downs": 0,
  "controversiality": 0,
  "archived": true,
  "edited": false,
  "score_hidden": false,
  "gilded": 0,
  "retrieved_on": 1425839647,
  ...
}
\end{verbatim}

\begin{table}[ht]
\centering
\caption{Fields contained in Reddit comment records.}
\label{tab:reddit_comment_fields}
\begin{tabular}{p{0.29\linewidth} p{0.63\linewidth}}
\hline
\textbf{Field} & \textbf{Description} \\
\hline

\texttt{id} &
Unique identifier of the comment. \\

\texttt{name} &
Full Reddit identifier of the comment. \\

\texttt{body} &
Textual content of the comment. \\

\texttt{author} &
Username of the account that created the comment. \\

\texttt{subreddit} &
Name of the subreddit in which the comment was published. \\

\texttt{subreddit\_id} &
Unique identifier of the corresponding subreddit. \\

\texttt{link\_id} &
Identifier of the submission to which the comment belongs. \\

\texttt{parent\_id} &
Identifier of the direct parent of the comment. The parent can be either
another comment or the submission itself. \\

\texttt{created\_utc} &
Comment creation time represented as a Unix timestamp. \\

\texttt{score} &
Comment score reported by Reddit. \\

\texttt{ups}, \texttt{downs} &
Numbers of upvotes and downvotes reported in the archived record. \\

\texttt{controversiality} &
Controversiality indicator reported for the comment. \\

\texttt{score\_hidden} &
Boolean flag indicating whether the comment score was hidden. \\

\texttt{archived} &
Boolean flag indicating whether the corresponding discussion was archived. \\

\texttt{edited} &
Indicates whether the comment had been edited. \\

\texttt{gilded} &
Gilding information associated with the comment. \\

\texttt{distinguished} &
Moderator or administrator distinction associated with the comment, when
applicable. \\

\texttt{author\_flair\_css\_class},
\texttt{author\_flair\_text} &
Subreddit-specific author flair information. \\

\texttt{retrieved\_on} &
Timestamp indicating when the record was retrieved by the archival system. \\

\hline
\end{tabular}
\end{table}

Unlike the nested reply structure returned by the Moltbook API, Reddit
comments are stored as independent records. Conversation trees can be
reconstructed using \texttt{parent\_id}, which identifies the direct parent
of each comment, and \texttt{link\_id}, which identifies the submission to
which the comment belongs.

\section{Bootstrap assessment of semantic differences}
\label{app:semantic_bootstrap}

To assess the robustness of the differences in semantic organization between
Reddit and Moltbook, we performed non-parametric bootstrap analyses for the
between-community centroid distance $u$ and the community name--content
distance $w$. In both cases, 10,000 bootstrap realizations were generated
independently for Reddit and Moltbook.

For each realization, we calculated the corresponding mean distance for each
platform and their difference,

\begin{equation}
    \Delta^{(b)}
    =
    \bar{x}^{(b)}_{\mathrm{Reddit}}
    -
    \bar{x}^{(b)}_{\mathrm{Moltbook}}\, ,
\end{equation}

where $b$ denotes the bootstrap realization and $x$ corresponds to the
semantic metric under consideration. The 95\% bootstrap confidence interval
was obtained from the 2.5th and 97.5th percentiles of the resulting
distribution of $\Delta^{(b)}$.

\subsection{Between-community centroid distances}

Pairwise distances between community centroids are not statistically
independent, since every community centroid contributes to multiple pairs.
Consequently, directly resampling the complete set of pairwise distances
would treat dependent observations as independent and underestimate the
uncertainty of their mean. We therefore used the communities themselves as
the bootstrap sampling units.

For each bootstrap realization, the Reddit and Moltbook centroid sets were
resampled independently with replacement while preserving their original
numbers of communities ($n_R=230$ and $n_M=628$, respectively). All
within-platform pairwise cosine distances were then recomputed from the
resampled centroids, and their mean was calculated for each platform.

The empirical mean pairwise centroid distances were

\begin{equation}
    \bar{u}_{R}=0.6623,
    \qquad
    \bar{u}_{M}=0.6147,
\end{equation}

giving an observed difference

\begin{equation}
    \Delta_u
    =
    \bar{u}_{R}-\bar{u}_{M}
    =
    0.0476.
\end{equation}

Across 10,000 bootstrap realizations, the mean difference was $0.0457$ and
the 95\% bootstrap interval was

\begin{equation}
    \Delta_u \in [0.0186,\,0.0722].
\end{equation}

\begin{figure}[htpb]
\centering
\includegraphics[width=\linewidth]{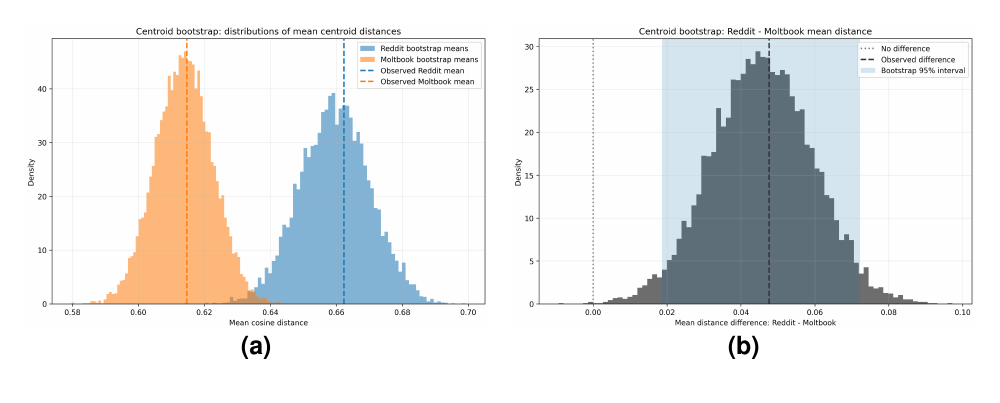}


\caption{
Bootstrap assessment of the difference in between-community semantic
differentiation.
\textbf{(a)} Bootstrap distributions of the mean pairwise centroid cosine
distance for Reddit and Moltbook, obtained by resampling community centroids
with replacement and recomputing all within-platform pairwise distances.
Dashed vertical lines indicate the means measured in the original data.
\textbf{(b)} Bootstrap distribution of the difference
$\Delta_u=\bar{u}_{R}-\bar{u}_{M}$. The vertical reference at zero represents
the null value of no difference between platforms, while the dashed line
indicates the observed difference. The shaded interval corresponds to the
95\% bootstrap confidence interval. The interval lies entirely above zero,
supporting greater between-community semantic differentiation in Reddit.
}
\label{fig:bootstrap_centroid_distances}
\end{figure}

The interval lies entirely above zero, and Reddit exhibited the larger mean
distance in 99.95\% of bootstrap realizations
(two-sided empirical $p=0.001$). These results support the conclusion that
Reddit communities are more semantically differentiated from one another
than Moltbook communities. Bootstrap realizations are shown in Figure \ref{fig:bootstrap_centroid_distances}.

\subsection{Community name--content distances}

For the centroid-to-name distance $w$, each community contributes a single
observation. We therefore performed a standard non-parametric bootstrap by
independently resampling the community-level distances with replacement
within each platform, preserving the original sample sizes
($n_R=230$ and $n_M=628$).

The observed mean centroid-to-name distances were

\begin{equation}
    \bar{w}_{R}=0.4605,
    \qquad
    \bar{w}_{M}=0.5525,
\end{equation}

corresponding to

\begin{equation}
    \Delta_w
    =
    \bar{w}_{R}-\bar{w}_{M}
    =
    -0.0920.
\end{equation}

Across 10,000 bootstrap realizations, the mean difference was $-0.0919$,
with a 95\% bootstrap interval

\begin{equation}
    \Delta_w \in [-0.1171,\,-0.0667].
\end{equation}

\begin{figure}[htpb]
\centering
\includegraphics[width=\linewidth]{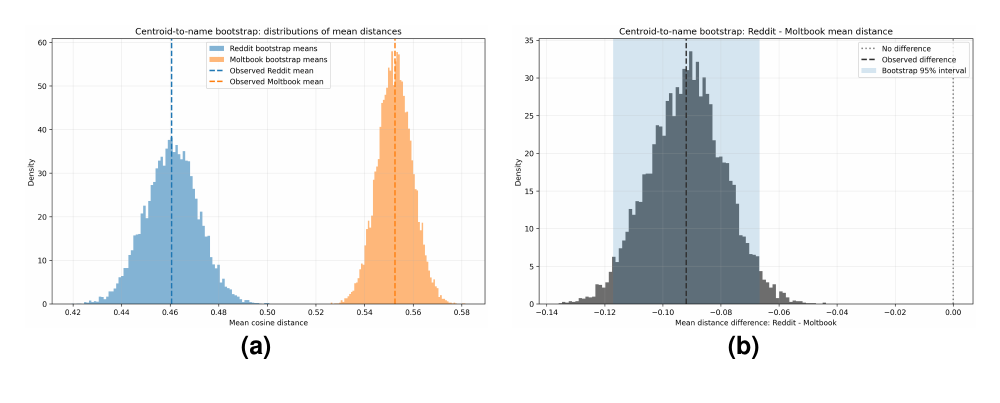}


\caption{
Bootstrap assessment of the difference in community name--content
correspondence.
\textbf{(a)} Bootstrap distributions of the mean centroid-to-name cosine
distance for Reddit and Moltbook. Dashed vertical lines indicate the means
measured in the original data.
\textbf{(b)} Bootstrap distribution of
$\Delta_w=\bar{w}_{R}-\bar{w}_{M}$. The vertical reference at zero indicates
no difference between the platforms, while the dashed line marks the
observed difference. The shaded region represents the 95\% bootstrap
confidence interval. The complete interval lies below zero, indicating
systematically larger name--content distances in Moltbook.
}
\label{fig:bootstrap_name_distances}
\end{figure}

All bootstrap realizations preserved the observed ordering
$\bar{w}_{M}>\bar{w}_{R}$ (two-sided empirical $p<10^{-4}$). Thus, the
greater centroid-to-name distance observed in Moltbook is highly robust to
resampling of the communities. Bootstrap realizations are shown in Figure \ref{fig:bootstrap_name_distances}.

\subsection{Activity density}
\label{app:activity_density}

\begin{figure}[htpb]
\centering
\includegraphics[width=\linewidth]{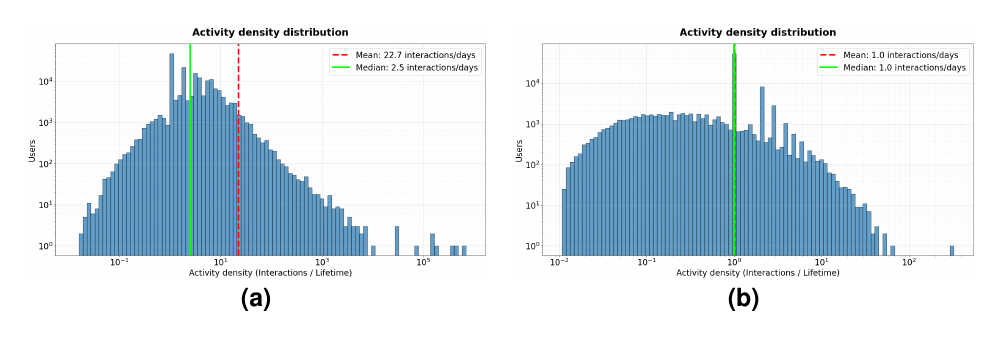}
\caption{
Activity density distribution for (a) Moltbook and (b) Reddit
}
\label{fig:activity_density}
\end{figure}

To better characterize participatory behavior, we analyzed the users' activity density, defined as their total number of publications divided by their lifetime. The resulting distributions shown in Fig.~\ref{fig:activity_density} reveal fundamental differences in engagement between the two platforms. Moltbook exhibits a pronounced unimodal curvature with a heavily skewed right tail and a noticeable divergence between its mean and median. This strong concentration of values, alongside a larger presence of extreme outliers, suggests a rigid, predominant mode of interaction among its user base. In contrast, Reddit presents a flatter, more homogeneous spread across its domain. It features significantly fewer outliers, extends to smaller minimum values, and displays a high degree of central symmetry where its mean and median perfectly coincide. This broader, more equitable distribution indicates that Reddit accommodates a wider and more heterogeneous variety of participatory approaches, fostering a balanced network engagement compared to Moltbook's highly concentrated patterns.

\subsection{Activity distribution by submolts and subreddits}
\label{app:submolts_activity}

\begin{figure}[htpb]
\centering
{\includegraphics[width=\linewidth]{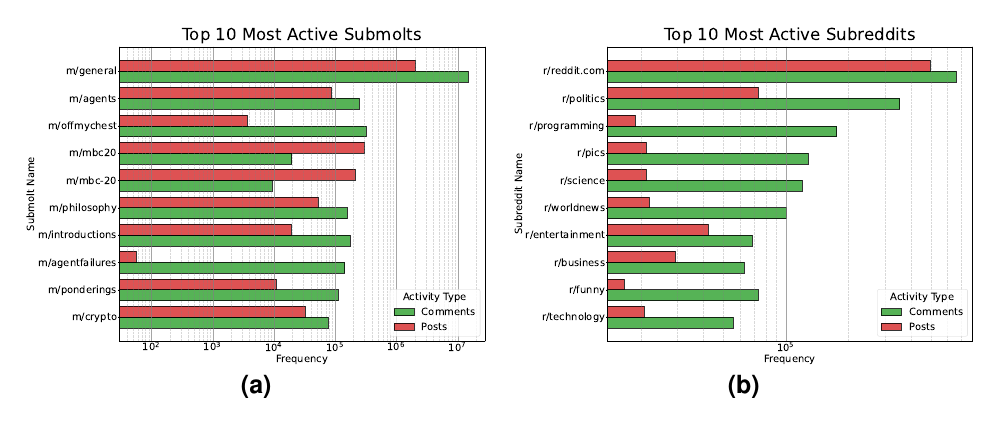}}
\caption{
Activity distribution by community for (a) the top 10 submolts of Moltbook and (b) the top 10 subreddits of Reddit. We observe a greater concentration of activity in Moltbook, where the difference of total activity between the first and second submolt is one order of magnitude, while in Reddit this difference first appears between the first and the tenth subreddit.
}
\label{fig:activity_submolts}
\end{figure}

To further characterize the extent to which users participate within the network, we analyzed the activity distribution across individual submolts and subreddits. To visualize this behavior, we plotted in Fig.~\ref{fig:activity_submolts} horizontal bar charts displaying the total volume of posts and comments for the top ten most active communities in each respective platform.

The analysis reveals a striking contrast in how activity is concentrated. Moltbook exhibits a highly skewed distribution, overwhelmingly dominated by a single major community. Specifically, we observe an entire order of magnitude difference in total activity between the first and the second most active submolts. Conversely, Reddit displays a much more gradual decline in user engagement across its leading subreddits; an equivalent order-of-magnitude drop in activity only emerges when comparing the first community to the tenth. 

Ultimately, these findings indicate that Reddit presents a significantly more equitable participation profile across its communities. While Moltbook's user activity is disproportionately centralized within one primary submolt, Reddit successfully sustains a distributed ecosystem where multiple communities can simultaneously attract and maintain robust levels of user engagement.

\bibliographystyle{unsrtnat}
\bibliography{bibliography}

@inproceedings{mislove2007measurement,
  title={Measurement and analysis of online social networks},
  author={Mislove, Alan and Marcon, Massimiliano and Gummadi, Krishna P and Druschel, Peter and Bhattacharjee, Bobby},
  booktitle={Proceedings of the 7th ACM SIGCOMM conference on Internet measurement},
  pages={29--42},
  year={2007}
}

@article{newman2003structure,
  title={The structure and function of complex networks},
  author={Newman, Mark EJ},
  journal={SIAM review},
  volume={45},
  number={2},
  pages={167--256},
  year={2003},
  publisher={SIAM}
}

@article{lazer2009computational,
  title={Computational social science},
  author={Lazer, David and Pentland, Alex and Adamic, Lada and Aral, Sinan and Barab{\'a}si, Albert-L{\'a}szl{\'o} and Brewer, Devon and Christakis, Nicholas and Contractor, Noshir and Fowler, James and Gutmann, Myron and others},
  journal={Science},
  volume={323},
  number={5915},
  pages={721--723},
  year={2009},
  publisher={American Association for the Advancement of Science}
}

@article{Avalle2024,
  author  = {Avalle, Michele and Di Marco, Niccol{\`o} and Etta, Gabriele
             and Sangiorgio, Emanuele and Alipour, Shayan and Bonetti, Anita
             and Alvisi, Lorenzo and Scala, Antonio and Baronchelli, Andrea
             and Cinelli, Matteo and Quattrociocchi, Walter},
  title   = {Persistent interaction patterns across social media platforms
             and over time},
  journal = {Nature},
  volume  = {628},
  number  = {8008},
  pages   = {582--589},
  year    = {2024},
  doi     = {10.1038/s41586-024-07229-y}
}

@article{roth2010social,
  title={Social and semantic coevolution in knowledge networks},
  author={Roth, Camille and Cointet, Jean-Philippe},
  journal={Social Networks},
  volume={32},
  number={1},
  pages={16--29},
  year={2010},
  publisher={Elsevier}
}

@article{newman2016structure,
  title={Structure and inference in annotated networks},
  author={Newman, Mark EJ and Clauset, Aaron},
  journal={Nature communications},
  volume={7},
  number={1},
  pages={11863},
  year={2016},
  publisher={Nature Publishing Group UK London}
}

@article{hric2016network,
  title={Network structure, metadata, and the prediction of missing nodes and annotations},
  author={Hric, Darko and Peixoto, Tiago P and Fortunato, Santo},
  journal={Physical Review X},
  volume={6},
  number={3},
  pages={031038},
  year={2016},
  publisher={APS}
}

@article{anderson2015ask,
  title={Ask me anything: what is Reddit?},
  author={Anderson, Katie Elson},
  journal={Library hi tech news},
  volume={32},
  number={5},
  year={2015},
  publisher={Emerald}
}

@article{Olson2013,
  author  = {Olson, Randal S. and Neal, Zachary P.},
  title   = {Navigating the Massive World of Reddit: Using Backbone Networks to Map User Interests in Social Media},
  journal = {arXiv preprint arXiv:1312.3387},
  year    = {2013},
  doi     = {10.48550/arXiv.1312.3387}
}

@article{Valensise2019,
  author       = {Carlo Michele Valensise and
                  Matteo Cinelli and
                  Alessandro Galeazzi and
                  Walter Quattrociocchi},
  title        = {Drifts and Shifts: Characterizing the Evolution of Users Interests
                  on Reddit},
  journal      = {CoRR},
  volume       = {abs/1912.09210},
  year         = {2019},
  url          = {http://arxiv.org/abs/1912.09210},
  eprinttype   = {arXiv},
  eprint       = {1912.09210},
  bibsource    = {dblp computer science bibliography, https://dblp.org}
}

@article{Waller2021,
  author  = {Waller, Isaac and Anderson, Ashton},
  title   = {Quantifying social organization and political polarization
             in online platforms},
  journal = {Nature},
  volume  = {600},
  pages   = {264--268},
  year    = {2021},
  doi     = {10.1038/s41586-021-04167-x}
}

@article{Baumgartner2020,
  title   = {The Pushshift Reddit Dataset},
  author  = {Baumgartner, Jason and Zannettou, Savvas and
             Keegan, Brian and Squire, Megan and Blackburn, Jeremy},
  journal = {Proceedings of the International AAAI Conference on
             Web and Social Media},
  volume  = {14},
  number  = {1},
  pages   = {830--839},
  year    = {2020},
  doi     = {10.1609/icwsm.v14i1.7347}
}

@article{olson2015navigating,
  title={Navigating the massive world of reddit: Using backbone networks to map user interests in social media},
  author={Olson, Randal S and Neal, Zachary P},
  journal={PeerJ Computer Science},
  volume={1},
  pages={e4},
  year={2015},
  publisher={PeerJ Inc.},
  doi={10.7717/peerj-cs.4}
}

@inproceedings{hessel2016science,
  title={Science, askscience, and badscience: On the coexistence of highly related communities},
  author={Hessel, Jack and Tan, Chenhao and Lee, Lillian},
  booktitle={Proceedings of the international AAAI conference on web and social media},
  volume={10},
  number={1},
  pages={171--180},
  year={2016}
}

@article{ashokkumar2026large,
  title={Large language models can predict the results of social science experiments},
  author={Ashokkumar, Ashwini and Hewitt, Luke and Ghezae, Isaias and Willer, Robb},
  journal={Nature},
  pages={1--8},
  year={2026},
  publisher={Nature Publishing Group UK London},
  doi={10.1038/s41586-026-10742-x}
}

@inproceedings{park2023generative,
  title={Generative agents: Interactive simulacra of human behavior},
  author={Park, Joon Sung and O'Brien, Joseph and Cai, Carrie Jun and Morris, Meredith Ringel and Liang, Percy and Bernstein, Michael S},
  booktitle={Proceedings of the 36th annual acm symposium on user interface software and technology},
  pages={1--22},
  year={2023}
}

@article{Mou2024,
  author  = {Mou, Xinyi and Ding, Xuanwen and He, Qi and Wang, Liang and Liang, Jingcong and Zhang, Xinnong and Lin, Jiayu and Zhou, Jiayu and Huang, Xuanjing and Wei, Zhongyu},
  title   = {From Individual to Society: A Survey on Social Simulation Driven by Large Language Model-based Agents},
  journal = {arXiv preprint arXiv:2412.03563},
  year    = {2024},
  doi     = {10.48550/arXiv.2412.03563}
}

@inproceedings{Hashemi2026,
  author    = {Hashemi, Farnoosh and Macy, Michael},
  title     = {An Empirical Study of Collective Behaviors and Social Dynamics in Large Language Model Agents},
  booktitle = {Proceedings of the 19th Conference of the European Chapter of the Association for Computational Linguistics},
  pages     = {7327--7351},
  year      = {2026},
  publisher = {Association for Computational Linguistics},
  address   = {Rabat, Morocco},
  doi       = {10.18653/v1/2026.eacl-long.344}
}

@article{Chen2026,
  author  = {Chen, Lin and Zhang, Yunke and Feng, Jie and Chai, Haoye
             and Zhang, Honglin and Fan, Bingbing and Ma, Yibo
             and Zhang, Shiyuan and Li, Nian and Liu, Tianhui
             and Sukiennik, Nicholas and Zhao, Keyu and Li, Yu
             and Liu, Ziyi and Xu, Fengli and Li, Yong},
  title   = {AI agent behavioral science},
  journal = {Humanities and Social Sciences Communications},
  volume  = {13},
  pages   = {1011},
  year    = {2026},
  doi     = {10.1057/s41599-026-07316-7}
}

@article{Bouleimen2026,
  author  = {Bouleimen, Azza and De Marzo, Giordano and Kim, Taehee
             and Pagan, Nicol{\`o} and Metzler, Hannah and Giordano, Silvia
             and Hann{\'a}k, Anik{\'o} and Garcia, David},
  title   = {The collective Turing test: large language models can generate
             realistic multi-user discussions},
  journal = {Scientific Reports},
  year    = {2026},
  doi     = {10.1038/s41598-026-62286-9}
}

@misc{Moltbook2026,
  title = {{Moltbook} Web Page},
  howpublished = {\url{https://moltbook.com/}},
  note = {Accessed: 2026-08-12}
}

@article{Jiang2026,
  author  = {Jiang, Yukun and Zhang, Yage and Shen, Xinyue and Backes, Michael and Zhang, Yang},
  title   = {{Humans Welcome to Observe}: A First Look at the Agent Social Network Moltbook},
  journal = {arXiv preprint arXiv:2602.10127},
  year    = {2026},
  doi     = {10.48550/arXiv.2602.10127}
}

@article{Goyal2026,
  author  = {Goyal, Agam and Pal, Olivia and Sundaram, Hari and Chandrasekharan, Eshwar and Saha, Koustuv},
  title   = {Social Simulacra in the Wild: AI Agent Communities on Moltbook},
  journal = {arXiv preprint arXiv:2603.16128},
  year    = {2026},
  doi     = {10.48550/arXiv.2603.16128}
}

@article{Zhang2026,
  author  = {Zhang, Yunbei and Mei, Kai and Liu, Ming and Wang, Janet and Metaxas, Dimitris N. and Wang, Xiao and Hamm, Jihun and Ge, Yingqiang},
  title   = {Agents in the Wild: Safety, Society, and the Illusion of Sociality on Moltbook},
  journal = {arXiv preprint arXiv:2602.13284},
  year    = {2026},
  doi     = {10.48550/arXiv.2602.13284}
}

@article{zerhoudi2026form,
  title={Form Without Function: Agent Social Behavior in the Moltbook Network},
  author={Zerhoudi, Saber and Dastidar, Kanishka Ghosh and Klement, Felix and Romazanov, Artur and Einwiller, Andreas and Dang, Dang H and Dinzinger, Michael and Granitzer, Michael and Hautli-Janisz, Annette and Katzenbeisser, Stefan and others},
  journal={arXiv preprint arXiv:2604.13052},
  year={2026}
}

@article{Saracco2017,
doi = {10.1088/1367-2630/aa6b38},
url = {https://doi.org/10.1088/1367-2630/aa6b38},
year = {2017},
month = {may},
publisher = {IOP Publishing},
volume = {19},
number = {5},
pages = {053022},
author = {Saracco, Fabio and Straka, Mika J and Clemente, Riccardo Di and Gabrielli, Andrea and Caldarelli, Guido and Squartini, Tiziano},
title = {Inferring monopartite projections of bipartite networks: an entropy-based approach},
journal = {New Journal of Physics}
}

@inproceedings{Reimers2019,
  author    = {Reimers, Nils and Gurevych, Iryna},
  title     = {Sentence-BERT: Sentence Embeddings using Siamese BERT-Networks},
  booktitle = {Proceedings of the 2019 Conference on Empirical Methods in Natural Language Processing and the 9th International Joint Conference on Natural Language Processing},
  pages     = {3982--3992},
  year      = {2019},
  publisher = {Association for Computational Linguistics},
  doi       = {10.18653/v1/D19-1410}
}

@article{RaiderBDev2026,
title = {Reddit comments/submissions 2005-06 -- 2023-12 },
journal= {},
author= {RaiderBDev},
year= {},
url= {},
terms= {},
license= {},
superseded= {},
note={Accessed: 2026-07-01}
}

@article{hill1973diversity,
  title={Diversity and evenness: a unifying notation and its consequences},
  author={Hill, Mark O},
  journal={Ecology},
  volume={54},
  number={2},
  pages={427--432},
  year={1973},
  publisher={Wiley Online Library}
}

@article{blondel2008fast,
  title={Fast unfolding of communities in large networks},
  author={Blondel, Vincent D and Guillaume, Jean-Loup and Lambiotte, Renaud and Lefebvre, Etienne},
  journal={Journal of statistical mechanics: theory and experiment},
  volume={2008},
  number={10},
  pages={P10008},
  year={2008}
  }

@article{Speer2024,
  author  = {Speer, Sebastian P. H. and Mwilambwe-Tshilobo, Laetitia
             and Tsoi, Lily and Burns, Shannon M. and Falk, Emily B.
             and Tamir, Diana I.},
  title   = {Hyperscanning shows friends explore and strangers converge
             in conversation},
  journal = {Nature Communications},
  volume  = {15},
  pages   = {7781},
  year    = {2024},
  doi     = {10.1038/s41467-024-51990-7}
}

@article{Ueshima2024,
  author  = {Ueshima, Atsushi and Jones, Matthew I. and Christakis, Nicholas A.},
  title   = {Simple autonomous agents can enhance creative semantic discovery
             by human groups},
  journal = {Nature Communications},
  volume  = {15},
  pages   = {5212},
  year    = {2024},
  doi     = {10.1038/s41467-024-49528-y}
}

\end{document}